# Effect of interface on mid-infrared photothermal response of MoS$_2$ thin film grown by pulsed laser deposition

Ankur Goswami[a*], Priyesh Dhandaria[a], Soupitak Pal[b], Faheem Khan[a],
Željka Antić[a], Ravi Gaikwad[a], Kovur Prashanthi[a], Thomas Thundat[a*]

[a]Department of Chemical and Materials Engineering, University of Alberta, Edmonton, Canada
[b]Department of Chemical Engineering, University of California, Santa Barbara, USA

**Abstract**

Here we report mid infrared (mid-IR) photothermal response of multi layer MoS$_2$ thin film grown on crystalline (p-type silicon and c-axis oriented single crystal sapphire) and amorphous substrates (Si/SiO$_2$ and Si/SiN) by pulsed laser deposition (PLD) technique. The photothermal response of the MoS$_2$ films was measured as changes in the resistance of MoS$_2$ films when irradiated with mid IR (7 to 8.2 μm) source. We show that it is possible to enhance the temperature coefficient of resistance (TCR) of the MoS$_2$ thin film by controlling the interface through proper choice of substrate and growth conditions. The thin films grown by PLD were characterized using XRD, Raman, AFM, XPS and TEM. High-resolution transmission electron microscopy (HRTEM) images show that the MoS$_2$ films grow on sapphire substrate in a layer-by-layer manner with misfit dislocations. Layer growth morphology is disrupted when grown on substrates with diamond cubic structure such as silicon due to growth twin formation. The growth morphology is very different on amorphous substrates such as Si/SiO$_2$ or Si/SiN. The MoS$_2$ film grown on silicon shows a very high TCR (-2.9% K$^{-1}$), mid IR sensitivity (ΔR/R=5.2 %) and responsivity (8.7 V/W) as compared to films on other substrates.

**Introduction**

Ultrathin 2D-layered transition metal dichalcogenides (TMDCs) with the formula of $MX_2$ (where M=Mo, W, Nb, Ta, Ti, Re and X=S, Se or Te) have attracted significant attention due to their potential applications in nanoelectronics[1, 2], optoelectronics[3, 4] valleytronics[5] and spintronics.[6] Among all the TMDCs, molybdenum disulphide ($MoS_2$) is a semiconductor ($\geq$1.9 eV) with direct electronic band gap when formed as a monolayer (~6.5 Å in thickness). However, it behaves like as an indirect band gap semiconductor with a band gap of 1.29 eV when the film thickness is more than 5 layers.

Devices based on monolayer $MoS_2$, specifically transistors exhibit high current density, excellent electrostatic integrity, large on/off ratio ($>10^8$), unprecedented carrier mobilities of 200-500 $cm^2$/ V s, and good electrical conductivity (~0.03 $\Omega^{-1} cm^{-1}$).[7, 8] In addition, single as well as multilayer (>5 layers) of $MoS_2$ have shown much promise as ultrasensitive visible and UV photodetectors[9], field emitter[10], gas sensors [11], piezoelectric and piezotronics devices[12]. The photodetection mechanism of $MoS_2$ depends on its high absorption in the UV-Vis range and generation of electron-hole pairs under photo-excitation.[9] This allows the device to produce large photocurrent/voltage under applied bias and reduces its electrical resistance. However, there are no such reports exist on variation of electrical resistance of $MoS_2$ under mid-IR range which can be the basis of mid IR detection devices.

For mid-IR (7 to 8.5 μm) detection, the materials require smaller band gap (140 to 170 meV) similar to that of routinely used HgCdTe (MCT) which works on photoconductive mechanism.[13] Despite outstanding performance as an IR detector, MCT suffers from the disadvantages such as weak Hg-Te bond, toxicity due to heavy metals and high power consumption.[14] In addition, this device requires cryogenic cooling in order to increase the high signal-to-noise ratio.[15] Hence, alternative materials such as vanadium oxide ($VO_2$) or amorphous silicon have been used as uncooled IR detectors and micro-bolometers which work on the resistance change due to IR illumination.[16-18] Despite much effort, limited number of literature is available on mid IR detection using these materials. However, a few recent reports on near IR and mid IR photothermal response of 2D materials such as graphene and graphene oxide (GO) appears promising.[19, 20] Recently, Bae et al. demonstrated photothermal effect of GO in the mid IR range (7 to 14 μm) under external heating.[19] Nevertheless, room temperature mid IR photothermal response of these materials is still lacking. However, because of the broadband mid IR absorption of $MoS_2$ it can show much enhanced photothermal response in mid IR range.[21] When exposed to mid IR, resistance of $MoS_2$ changes due to photothermal effect. Resistance variation of a material due to heating depends on its temperature co-efficient of resistance (TCR). Hence, materials with high TCR can be used for IR detection using photothermal effect. It has been reported that the TCR of a thin film greatly influenced by its microstructure and the substrate-film interface for materials such as $La_{0.7}Ca_{0.3}MnO_3$ on $LaAlO_3$ and $SrTiO_3$ substrates.[22] However, there is no systematic study on the effect of $MoS_2$ film-substrate interface and film morphology on the TCR exist for $MoS_2$ films except a brief report in literature.[23]

In this work, we have systematically investigated the growth condition and behaviour of $MoS_2$ thin films on four different substrates, deposited by pulsed laser deposition (PLD). We have also explored the photothermal response of the $MoS_2$ films in the mid IR range (7 to 8.2 μm). The PLD technique offers great advantages as it directly transfers materials from the target to the substrate rapidly to achieve uniform deposition. However, it also produce defects on substrate-film interface due to highly non-equilibrium growth and the high ion bombardment from the target to the substrate.[24, 25] This can attribute remarkable property changes in the film.[26] For instance, in terms of electrical properties of $MoS_2$, the defects can introduce additional energy states in the band gap, which can change the n-type $MoS_2$ thin film to p-type and vice versa.[27] Introduction of phonon assisted phenomenon due to the presence of defect states in the band gap has been demonstrated by Raman spectroscopy.[28] Hence, two crystalline substrates of different crystal structures (e.g. p-type silicon and single crystal sapphire) as well as two amorphous substrates (e.g. $Si/SiO_2$ and Si/SiN) with no specific orientations have been chosen for this study. Though there are few reports on substrate effect on the optical and electrical properties of $MoS_2$, they are restricted to film transfer method only.[29, 30] Therefore, it is possible to enhance TCR of $MoS_2$ films by choosing the right substrate for deposition. In order to optimize the structural characteristic of the film, we have employed the simple concept of theoretical lattice mismatch strain between the highest atomic density plane of $MoS_2$: (0001) and planes parallel to the substrate surface. The hcp crystal structure of $MoS_2$ and (0001) orientation of sapphire allow a theoretical lattice mismatch strain of ~ 6% that result in semi-coherent interface. On the other hand, the theoretical lattice mismatch strain between p-type silicon and the highest atomic density plane of $MoS_2$ is ~ 40% and can form an incoherent interface. Meanwhile the film grown on amorphous substrates having no specific orientation can grow into large structures since there is no lattice mismatch strain in the film and the substrate. Therefore, by controlling the lattice mismatch strain between substrate and the $MoS_2$ film it is possible to tune the TCR characteristics and mid IR response.

**Experimental**

**Materials synthesis and deposition**

A two-inch $MoS_2$ target was prepared by pelletizing $MoS_2$ powder (purity 99% from Sigma Aldrich) at a load of 50 kN force using hydraulic press. The pellet was sintered at 800 °C for 12 h by purging Ar gas in a tubular furnace. The MoS$_2$ pellet was then mounted on a target holder inside the PLD chamber (Excel instruments, Mumbai, India). Four different substrates were used for deposition: p-type silicon <100> oriented ($\rho$=10-20 $\Omega$-cm), single crystal sapphire (0001), thermally grown silicon oxide (amorphous SiO$_2$~500 nm) and low pressure chemical vapour deposition (LPCVD) grown low stress silicon nitride (amorphous SiN~250 nm) on silicon. The deposition was carried out at two different substrate temperatures of 700 °C to 800 °C to investigate the effect of deposition temperature on growth morphology. The distance between the target and substrate was maintained at 5 cm for all the depositions. The base vacuum of the chamber was maintained (1 to 2) ×10$^{-5}$ torr. The deposition was done in the presence of argon (Ar) at a chamber pressure of 1.1 ×10$^{-2}$ torr. Krypton fluoride (KrF, $\lambda$=248 nm) excimer laser (Coherent, GmbH) was used with a 20 ns pulse width with a repetition rate of 5 Hz. Two different laser energies (35 and 50 mJ) were used for deposition. The effective laser fluence was approximately 1.2 J/cm$^2$ (energy: 35 mJ) and 1.7 J/cm$^2$ (energy: 50mJ) considering laser spot size of 3 mm × 1 mm measured at the target. The deposition time was varied from 5 s to 300 s to investigate its effect on morphology of MoS$_2$ films. The substrates were cooled down to room temperature at the same Ar atmosphere while maintaining the chamber pressure (i.e. 1.1×10$^{-2}$ torr) constant. The different deposition parameters that are used for growing MoS$_2$ thin films on various substrates are listed in Table-S1 (see ESI).

**Characterization**

The PLD deposited samples were characterized by X-ray diffraction (Cu K$_\alpha$) using Rigaku XRD Ultima 4 at glancing angle mode with incident angle of 0.5°. The Raman spectroscopy was done by Almega XR dispersive Raman microscope (Nicolet, Thermo Scientific) at 5 mW laser power using 50 × objectives. Excitation wavelength of 532 nm was used and a spot size of 1 μm was maintained in order to avoid possible heating effects. The surface topography and the thickness of the films were measured by Dimension Fast Scan Atomic Force Microscope (Bruker Nanoscience division, Santa Barbara, CA, USA). Commercially available Pt-Ir coated conductive probes (SCM-PIT) with a spring constant of 2.5 N/m and a resonant frequency of 65 kHz were used for obtaining surface topography. SEM imaging was done using Zeiss Sigma. X-ray photoelectron spectroscopy (XPS) was performed at a base vacuum of $1.5 \times 10^{-8}$ torr using Kratos imaging spectrometer to discern the chemical composition of the film. UV-Vis spectroscopy of the $MoS_2$ samples (deposited on sapphire) was carried out using Perkin Elmer spectrophotometer and the photoluminescence was done using LabRAM HR system. The film thickness and morphology were characterized using SEM (Zeiss Sigma) and TEM (Titan, FEI, the Netherlands operated at 300 kV). TEM foils of less than 100 nm thicknesses were prepared using Focused Ion Beam (FIB) machine (Hellios 600, FEI, The Netherlands) followed by lift off method. Film thickness measurements and diffraction analyses were carried out to determine the crystallinity through the Fast Fourier Transform (FFT) method using Digital Micrograph Software (Gatan Inc.).

**Mid IR photothermal Response and Electrical Characterization**

Mid IR photothermal IR characterization of deposited $MoS_2$ films was carried out using a quantum cascade laser (QCL) model ÜT8, Daylight Solutions, USA. The QCL was operated at 5% duty cycle pulsed at 100 kHz with a peak power of 400 mW in the mid-IR range (1200 $cm^{-1}$ to 1400 $cm^{-1}$ ~ 8.3 to 7.1 μm) and with a spot size of ~2.5 mm. In these experiments average power of the QCL was varied up to 25 mW. The contact pads of Ti/Au (5/50 nm) of 0.5 mm diameter separated by 1 mm distance were deposited by e-beam evaporation technique using aluminium hard mask. The films were annealed at 200 °C for 1 h in vacuum (10 mtorr) oven in order to reduce the contact resistance. Relatively thicker films (300 s deposited at 800 °C) of $MoS_2$ deposited on different substrates were used for this study in order to ensure a measurable resistance within the contact pad distance. Electrical resistance of the films were measured by two-probe method using Keithley 194 digital multimeter using LabView interface. Change of resistance was monitored by electrically pulsing the laser for every 120 s without any external bias. Temperature dependent resistance measurements were carried out using Signatone probe station 1160 series on a heating chuck where temperature was varied from 23 to 110 °C at an interval of 2 °C with an equilibration time of 5 min at each interval.

**Results and Discussion**

$MoS_2$ was deposited on various substrates at two different temperatures and deposition time as mentioned earlier. Detailed structural and morphological characterization of the samples deposited at 700 °C is discussed in ESI. Fig. 1 shows the XRD and Raman characterization of as deposited $MoS_2$ thin film grown on various substrates at 800 °C. Thin films of $MoS_2$ deposited at 800 °C show higher degree of crystallinity for $MoS_2$ than the one deposited at 700 °C. Fig.1a shows that film crystallinity improves as the time of deposition increases to 300 s.

Raman spectroscopy was performed on all the samples to confirm the formation of $MoS_2$. From Raman spectra it is evident that the samples deposited at 800 °C temperatures show appreciable crystallinity for 20 s as well as 300 s deposition time. It is discernible from Fig. 1b that the layer thickness of the $MoS_2$ for 20 s deposition times is around 3 to 4 layers (see Table S1 in ESI). Fig. 1c shows the formation of an ultrathin $MoS_2$ observed at lower deposition time (5 s). Except for the silicon, all the other substrates show appreciable growth of 3~4 layers at that deposition condition. Fig. 1d shows the photograph of MoS$_2$ grown on the sapphire substrates at 800 °C for two different deposition times and energy. This indicates uniform coverage of the film and the changes in the optical transparency of sapphire with increasing thickness of MoS$_2$ layers.

Surface morphologies for the films grown at 800 °C deposition temperature for 20 s of deposition time at 35 mJ laser energy are different for all the substrates as shown in the AFM image in Fig.2 (a) and Fig.S3. The growth of MoS$_2$ on silicon is disrupted since (0001) plane of MoS$_2$ grows on Si (100) which is not lowest surface energy plane of silicon. Consequently, adatom of the substrate surface required to overcome the crystalline barrier of (100) plane of silicon which creates more strain in the film resulting hairline streaks on the surface (see Fig.S3a). However, MoS$_2$ grown on sapphire show a triangular morphology (as shown in Fig.2a). This could be possibly due to the growth of MoS$_2$ on the basal plane (0001) of the sapphire which is a hexagonal close packed structure (*hcp*) similar to 2H-MoS$_2$ structure. On the other hand, morphology of MoS$_2$ deposited on thermally grown oxide (Si/SiO$_2$) resembles a sheet with thickness of 1.5 nm (as shown in Fig. S2b). This indicates that 2 ML of MoS$_2$ sheet formed on the Si/SiO$_2$ substrates. The average surface roughness of MoS$_2$ deposited on all the substrates at 800 °C for 20 s is in the range of 0.3 to 0.6 nm. Further, at 800 °C deposition temperature and 300 s of deposition time, the morphology shows a Stransky-Karstinov type growth. This results in the formation of dense nano structures of MoS$_2$ in all the substrates as shown by SEM in Fig.S4(a & b) (see ESI) and agrees with the findings of Late et al.[10] The thickness of the film (300 s deposited) found from cross sectional SEM was 16 to 18 nm as shown in Fig.S4c (see ESI). Since the samples deposited at 800 °C are structurally optimized, further studies on photothermal response were carried out on these samples.

X-ray photoelectron spectroscopy (XPS) was carried out for all the thin films grown at 800 °C and 20 s deposition times and shown in Fig.S5 and S6 (see ESI). All the films deposited on different substrates show typical binding of Mo-3d and S-2p which confirms the typical $MoS_2$ growth at this condition. A slight peak of $Mo^{6+}$ indicates possibility of $MoO_2$ and $MoO_3$ existence. However, there is no amorphous sulphur in the peak indicating the samples are of highly crystalline quality.

The optical properties of the films were characterized by UV-Vis and photoluminescence spectra and shown in Fig. S7(see ESI). Detailed analysis and the data are presented in the ESI. However, from the PL spectra (Fig.S6b, see ESI) it is clear that SAP8-5 sample shows a peak at 653 nm (1.89 eV) which is the direct excitonic transition from the band gap since this contains 2 to 3 layers of $MoS_2$ as also confirmed from Raman (TableS2, see ESI) and AFM measurements. As the number of layers increases the peak show red shift and while the peak intensity reduces. For bulk $MoS_2$ (SAP8-300) the signal intensity is very small as reported in earlier literature.[31] This confirms the band gap of $MoS_2$ changes with layer numbers and the band gap of the multilayer $MoS_2$ deposited here in all the substrates converges to 1.29 eV as reported in various literature.[32, 33]

The growth morphology and film substrate interface were characterized using transmission electron microscope (TEM). The HR-TEM image of the SAP8-20 specimen (see Fig. 3a) shows staking of $MoS_2$ layer on sapphire substrate, whereas FFT pattern (see Fig. 3b) captured from the film substrate conjugate suggest the foil normal is $<11\bar{2}0>$. The spots corresponding to the (0002) planes of $MoS_2$ with d-spacing of 6.147 Å have been identified in the FFT image and marked using yellow arrows. Other spots from the $MoS_2$ film have not been detected in the FFT pattern due to the presence of few atomic layers in the film which reduces the amplitude of the exit wave function. Careful observation of HRTEM image of Fig. 3a shows that the film formation occurs through stacking of a few layers (4-5 ML) of $MoS_2$ by forming a coherent relationship with the sapphire substrate. A schematic of the observed orientation relation between the sapphire and $MoS_2$ film is shown in Fig.S8a (see ESI), where the orientation relationship follows $(0003)_{Al2O3}\|(0002)_{MoS2}$ and $<0001>_{Al2O3}\|<0001>_{MoS2}$. Eventually, $<11\bar{2}0>$ direction is perpendicular to the <0001> direction and XRD also shows a strong (0002) peak of $MoS_2$. These altogether confirm that growth of the $MoS_2$ film on sapphire substrate occurs through formation of layered structured film. HRTEM observation also confirms to the layer numbers determined from Raman spectra as mentioned in Table S2 (see ESI) and AFM image shown in Fig.2a.

The interface formation between the silicon and $MoS_2$ is different from that of $MoS_2$-sapphire interface. HRTEM image of the film-substrate interface of $MoS_2$-Si shows that (020) types of planes of silicon are parallel to the interface, whereas the foil normal is $<10\bar{3}>$ direction as shown in Fig.3c. In a similar fashion $MoS_2$-sapphire interface, (002) planes of $MoS_2$ are parallel to (003) planes of sapphire and the interface exhibits relatively smaller lattice mismatch strain of ~ 5.6 %. As a result, it can accommodate the strain by forming semi-coherent interface creating subtle misfit dislocations as shown in Fig. 3(a) inset. For the $MoS_2$-silicon system, the theoretical lattice mismatch between Si(020) plane and $MoS_2$ (0002) plane is approximately -42%. Therefore, formation of the $MoS_2$ layer while maintaining this huge strain misfit is not practically possible. In order to minimize the strain energy, the film grown on the silicon substrate forms an incoherent interface through twin formation as shown in Fig. 3(d). This phenomenon is common in the case of *hcp* crystal system when growth occurs on diamond cubic or *fcc* structure through non-equilibrium processes like PLD.[34] Twining formation in non-equilibrium growth of *hcp* films occurs due to its low stacking fault energy and limited availability of slip planes.[35]

The mid-IR photothermal response of MoS$_2$ was studied by illuminating the top surface of the MoS$_2$ films grown on four different substrates using QCL. In the following text samples are named by the sample code mentioned in Table S1 i.e. S8-300, SAP8-300, SO8-300 and SN8-300. These samples are relatively higher in thickness (16 to 18 nm) than the 20 s deposited (thickness~ 3 to 4 nm) samples. However, it is obvious that the film substrate interface of these samples (300 s) would be similar to the 20 s deposited samples. Hence the explanation related to the film-substrate interface of 300 s deposited samples are based on the TEM studies of 20 s deposited one. Fig.4a and 4b show the schematic and the photograph of the experimental set up of the MoS$_2$ thin film under IR illumination using a QCL. A baseline of the photo-response was taken on the bare substrates that showed no significant change in the resistance upon IR illumination. Since, most of the substrates used in this work (except silicon) are highly insulating at room temperature (resistivity~ $10^{14}$ ohms-cm)[36-38] the resistance of the bare substrates fluctuates a large degree with no response to IR on/off pulses (data not shown). Fig. 4c shows the variation of resistance of MoS$_2$ on sapphire as a function of different wavelengths of pulsed IR. Rise ($\tau_{rise}$) and fall ($\tau_{fall}$) time of the device (SAP8-300) to reach 63 % of the saturation state is found to be 9 s and 10 s respectively as shown in Fig.4d. The IR data of the other samples (i.e. S8-300, SO8-300 and SN8-300) are shown in Fig.S9 (see ESI). The time constant is calculated by fitting an exponential decay/rise function of the measured data.

Fig. 5a shows the inter-comparison of the device sensitivity in terms of resistance change (i.e. sensitivity=$\frac{\Delta R}{R_0} \times 100\%$, where $\Delta R = R_0 - R$, $R_0$ and $R$ are resistance before and after IR illumination respectively). MoS$_2$ on silicon (S8-300) shows significantly higher sensitivity than any other devices.

The observation of mid-IR response of MoS$_2$ thin film is quite interesting. The band gap of the MoS$_2$ used for the mid-IR measurement was 1.29 eV since the films are multilayered and they are considered to have bulk properties. The energy of photons in the mid-IR range (7 to 8.2 µm) falls within the range of 140 to 170 meV and is much lower than the band gap energy of bulk MoS$_2$. We anticipate that there are two mechanisms that dominate the mid-IR response. First and the foremost mechanism involve the much-enhanced IR absorption (from FTIR data) of MoS$_2$ in this wavelength range (6 to 9 µm) as shown in Fig. S10 (see ESI). This observation is similar to the photothermal response of 2D thin films of graphene[20], graphene oxide[39] and 1-D nanowires.[40]

As mentioned earlier, the temperature coefficient of resistance, TCR $\left(\alpha = \frac{1}{R_0}\frac{dR}{dT}\right)$, is the important parameter to estimate the mid-IR response of MoS$_2$. Fig.5 (b and c) show the temperature dependence of relative resistance and the TCR of the MoS$_2$ film on all the substrates due to the externally applied heat. On an average, the TCR of MoS$_2$ is found to vary from -0.9% K$^{-1}$ (23 °C) to -0.3 % K$^{-1}$ (110 °C) for most of the substrates (sapphire, thermal oxide and SiN) which is commendable for 2D material in comparison to graphene and strongly reduced graphene oxide (s-GO). [19] However, there is a strong substrate dependence of TCR of MoS$_2$ as seen in case of the film grown on silicon. The TCR is found to vary from -2.9 % K$^{-1}$ (23 °C) to -0.3 % K$^{-1}$ (110°C) in S8-300 sample as shown in Fig. 5c. This could be due to the twin formation between silicon and MoS$_2$ interface as discussed earlier section on TEM studies. The twin boundaries play an important role in the electrical and thermal conductivity since they are weak scattering centre of electron and phonon.[34, 41, 42] As a result, increasing temperature increases the mobility of electrons due to change in the effective mass resulting in higher TCR for MoS$_2$ on silicon.[34, 41] Similar observation was also made by Zande et.al where in plane electrical conductivity of $MoS_2$ was found to increase due to mirror twin boundaries.[43] It is also possible that twin boundaries at the interface may open up interfacial defect conduction which increases with temperature resulting in higher TCR. On the other hand, as discussed earlier sapphire and $MoS_2$ has a lattice mismatch of 5.6% and forms a semi-coherent interface having low elastic strain (as shown in Fig.3a) with subtle misfit dislocations. Similarly, low strain interface also forms when the film grows on amorphous substrates such as thermally grown oxide or SiN because of the absence of long range lattice ordering in the substrates. Hence, these films are also prone to form dislocations at the interface. Dislocations are the coulomb scattering centres for electron pathways which significantly reduce the electron mobility in $MoS_2$ and 2D electron gas.[44-46] Therefore, $MoS_2$ grown on all the three substrates (i.e. sapphire, thermally grown oxide and SiN) show relatively low TCR values due to increased number of scattering centres. Hence, their TCR remains in the similar range as depicted in Fig.5b. The $MoS_2$ grown on silicon shows higher photothermal sensitivity than $MoS_2$ grown on any the other substrates due to its high TCR. However, this observation demands thorough theoretical understanding of phonon and electron transport across the $MoS_2$-substrate interface with increasing temperature.

Apart from the TCR, there is also another possibility which can contribute slightly to the mid-IR response. The XPS analyses (Fig.S6) show that oxygen molecules get adsorb on the $MoS_2$ surface after the film deposition. This can cause the multiple surface traps at the interface. These surface traps in $MoS_2$ are generally within the vicinity of ~200 meV of VB and CB as described by Tongay et al.[47] As a result, low energy IR would be enough to excite more electrons to the CB and thereby reducing the overall resistance. Though this phenomenon justifies overall IR response of MoS$_2$ on all the substrates, it does not explain the enhanced response of MoS$_2$ on silicon.

In order to understand the IR response of the film a detailed figure of merit calculation and experimental measurements were carried out as tabulated in Table I. Infrared responsivity $\left(R_{Resp} = \dfrac{V}{W}\right)$, thermal noise (or Johnson noise) $\left(V_n = \sqrt{4k_B TR\Delta f}\right)$, (where $k_B$ is the Boltzmann constant, $\Delta f$ is the bandwidth of measurement), noise equivalent power $\left(NEP = \dfrac{V_n}{R_{resp}}\right)$ and detectivity $\left(D^* = \dfrac{R_{resp}\sqrt{A}}{NEP}\right)$, ($A = 5\,mm^2$) of MoS$_2$ on different substrates are reported. Definition of all these figures of merit can be found elsewhere.[19, 48] It is noted from Table I that MoS$_2$ on silicon (S8-300) shows highest responsivity than any other substrates investigated here. However, it shows relatively high thermal noise due to higher electrical resistance of the film at room temperature. In case of thermal detectors, the response time is an important parameter which largely depends on the thermal characteristics such as thermal diffusivity. Substrates having high thermal diffusivity show higher response and recovery time than the other substrates. This is evident for MoS$_2$ films deposited on silicon (S8-300) as shown in Table I. However, in all these cases the MoS$_2$ films were on the substrates hence the characteristic times were higher due to the higher thermal mass of the system. Using suspended MoS2 structures could reduce the effect of thermal mass, and that could result in a highly responsive and sensitive bolometer.

**Conclusions**

In summary, MoS$_2$ thin films were grown on both crystalline and amorphous substrates using PLD technique. The deposition process was optimized for substrate temperature, deposition time and energy of the laser. Thin films of MoS$_2$ consisting of few layers were grown uniformly on various substrates and characterized by different techniques such as XRD, Raman, AFM, SEM, TEM and XPS. The results show that MoS2 growth is optimized at 800 °C with growth morphology showing clear dependence on substrate type. The TEM results show that the growth of the MoS$_2$ on silicon proceeds through twinning due to incoherent interface formation whereas the MoS$_2$ on sapphire forms layer-by-layer structure through subtle misfit dislocations. The resistance of the MoS$_2$ film show strong mid IR responsivity due to the broadband mid IR absorption. It was observed that the MoS$_2$ films grown on silicon offer much higher IR sensitivity and responsivity than the other substrates. This result can be explained as due to the high TCR stemming from the twin boundary formation caused by the large lattice mismatch strain between silicon and MoS$_2$. Therefore controlling the interfacial strain of the MoS$_2$ film by proper choice of substrate offers a way for enhancing its mid IR responsivity.

**Acknowledgement**

This work was supported by the Canada Excellence Research Chair (CERC) program. The authors also acknowledge the characterization facilities provided by Alberta Centre for Surface Engineering & Sciences (ACSES), Oil Sands & Coal Interfacial Engineering Facility (OSCIEF) and the Nanofab at the University of Alberta. Authors thank Mr. Richard Hull for useful discussions on noise analysis.

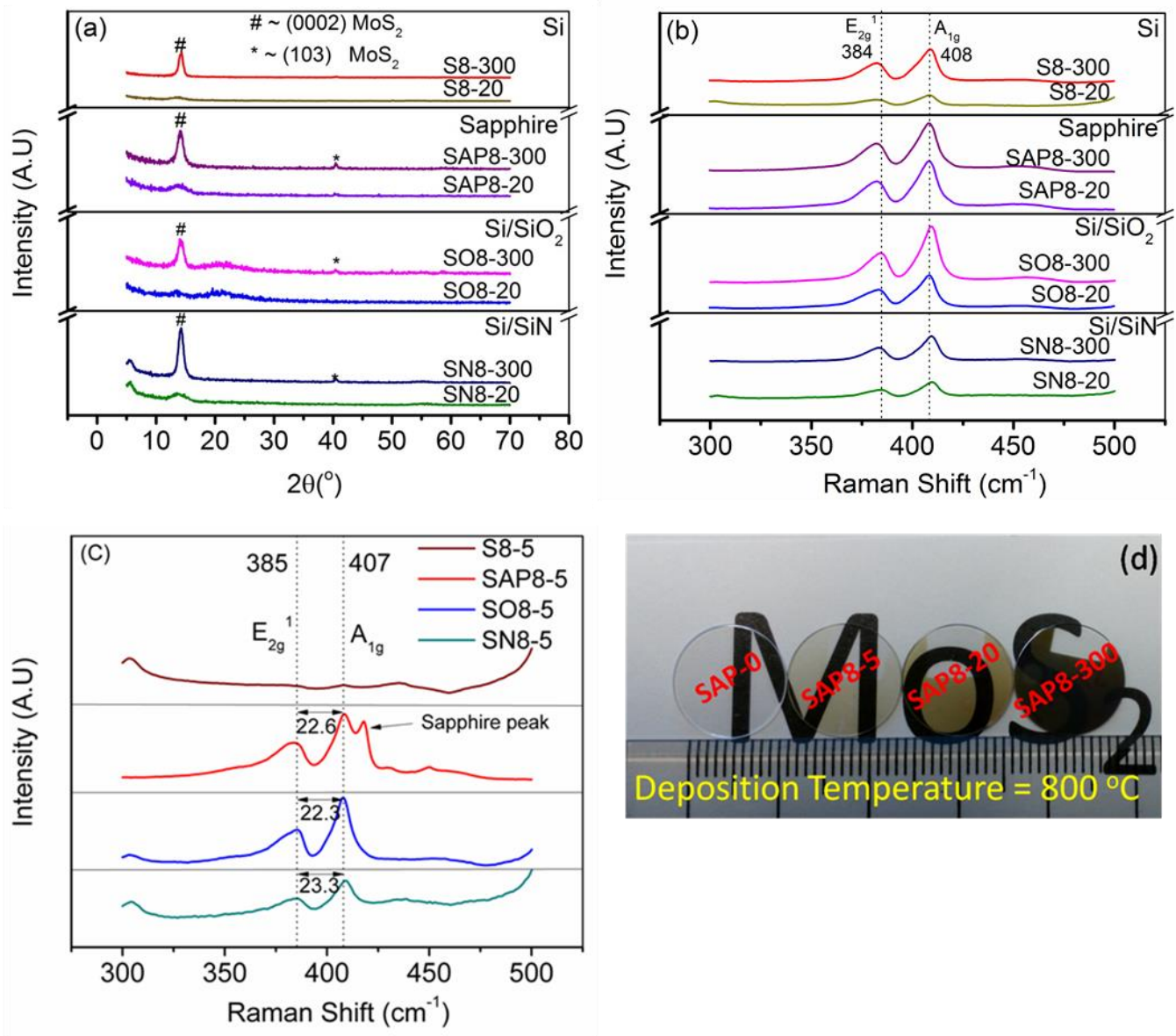

**Fig.1:** (a) XRD of MoS$_2$ grown on different substrates at 800 °C and 35 mJ energy. Raman spectra of MoS$_2$ grown on different substrates at 800 °C at (b) 35 mJ and (c) 50 mJ laser energy respectively for different times of deposition. (d) Photograph of the MoS$_2$ film grown on sapphire at 800 °C different time and laser energy. The word MoS$_2$ was printed on the background of the substrate.

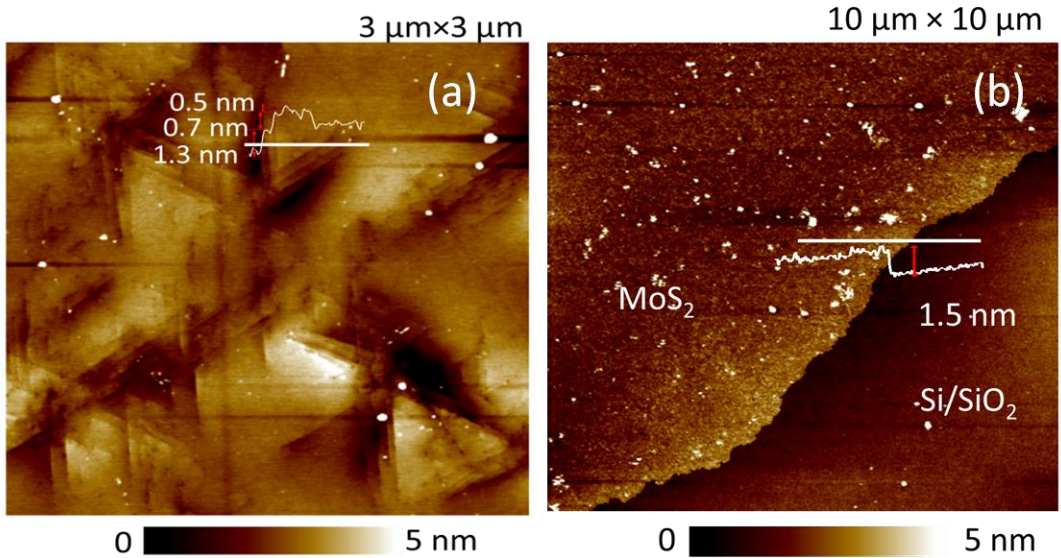

**Fig. 2:** AFM images of MoS$_2$ grown at 800 °C on (a) sapphire (SAP8-20) at 35 mJ 20 s, (b) thermally grown oxide (SO8-5) at 50 mJ for 5 s. The uncovered area of thermally grown oxide in Fig.2(b) is because hard mask was on the substrate while depositing MoS$_2$.

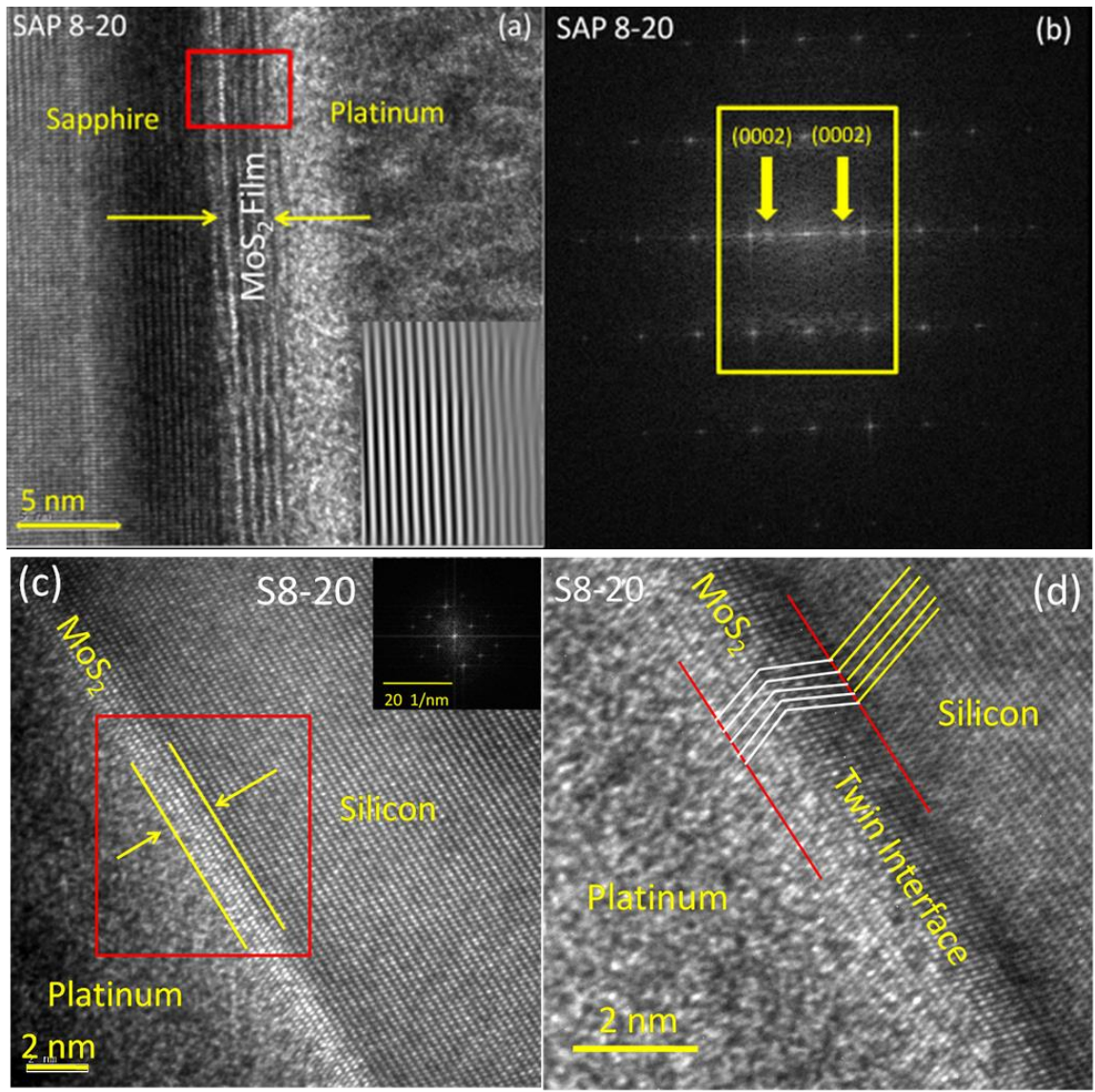

**Fig.3 :** (a) Cross-sectional TEM image of MoS$_2$ film grown on sapphire (0001) at 800 °C 20 s deposition time showing stacking of MoS$_2$ with (0002) orientation. Inset shows the inverse FFT image of the red mark area of the image showing misfit dislocations. (b) FFT pattern of the corresponding red mark zone of image (a). The orientation relationship of the yellow mark area of the FFT pattern depicted in Fig.S4 (see ESI). (c) Cross-sectional TEM image of MoS$_2$ grown on silicon <100>. Inset shows the FFT image of the image showing no orientation relationship. (d) Twin formation of silicon- MoS$_2$ interface.

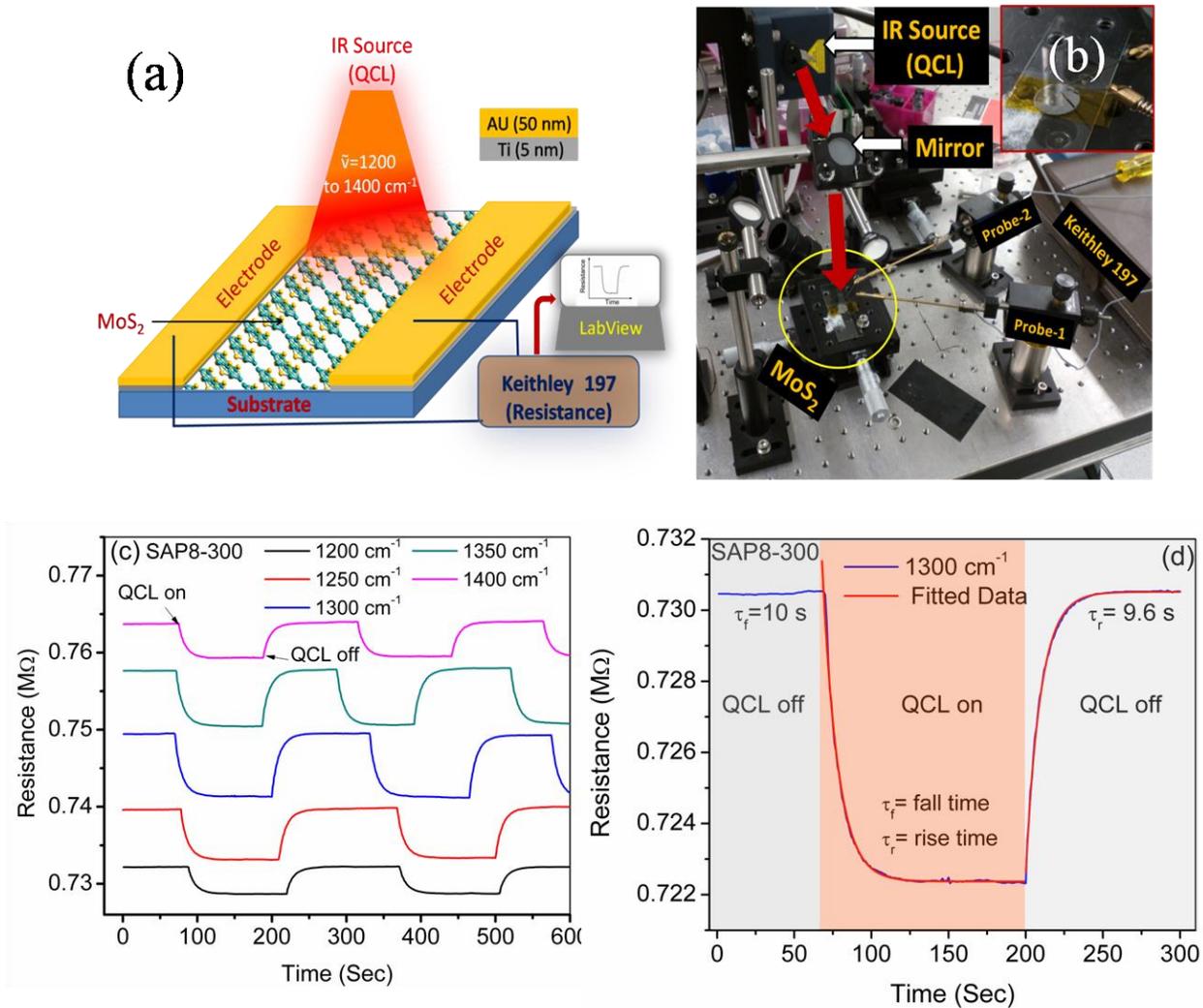

**Fig.4** (a) Schematic and (b) photograph of the experimental set up of mid-IR response of MoS$_2$ on different substrates. (c) Variation of resistance under mid-IR illumination of MoS$_2$ on sapphire substrate (SAP8-300) at different wave numbers. (The data was plotted using 2D waterfall mode at 10% offset in order to accommodate all the data set). (d) Shows the photothermal response time of MoS$_2$ on sapphire (SAP8-300).

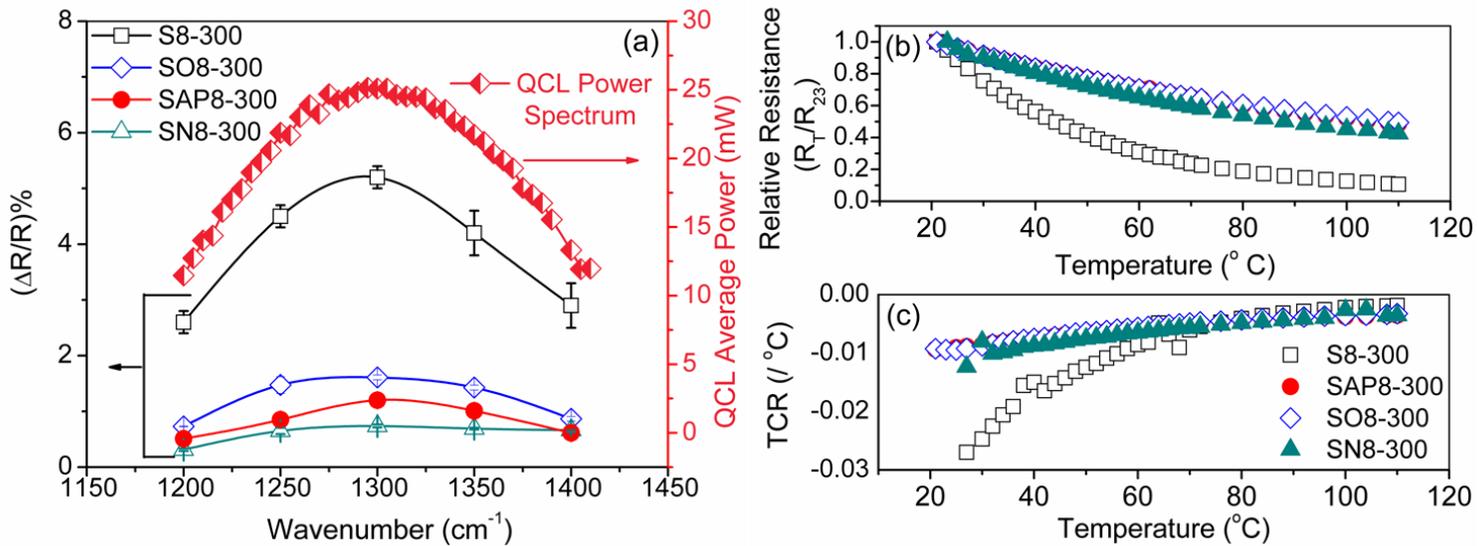

**Fig.5 :** (a) Sensitivity of photothermal response of $MoS_2$ grown on different substrates. In the same graph power spectrum of the QCL is shown. (b) Relative resistance change and (c) TCR of $MoS_2$ on different substrates with function of temperature using external heating source from the bottom of the substrate. The data is normalized with the resistance at T=23 °C.

Table I: Estimated mid-IR response characteristics of MoS$_2$ on different substrates at 1300 cm$^{-1}$ (7.7 μm) at its highest average power (25 mW). IR exposure area (A) was 5 mm$^2$. Johnson noise is reported per unit bandwidth (i.e. $\Delta f = 1$ Hz).

| Sample | Substrates | Thermal diffusivity of the substrates (cm$^2$/s) | Responsivity ($R_{resp}$) (V/W) | Johnson Noise or Thermal Noise ($V_n$) (V Hz$^{-1/2}$) | Noise Equivalent Power (NEP) (W Hz$^{-1/2}$) | Response time (sec) $\tau_f$ & $\tau_r$ | Detectivity (D*) (cm Hz$^{1/2}$ W$^{-1}$) |
|---|---|---|---|---|---|---|---|
| S8-300 | Si | 800×10$^{-3}$ | 8.7 | 5.9×10$^{-7}$ | 6.8×10$^{-8}$ | 8.5 & 7.5 | 2.8×10$^7$ |
| SAP8-300 | Sapphire | 83×10$^{-3}$ | 1.9 | 1.1×10$^{-7}$ | 5.8×10$^{-7}$ | 10 & 9.6 | 7.2×10$^6$ |
| SO8-300 | Si/SiO$_2$ | 6.4×10$^{-3}$ | 2.6 | 4.7×10$^{-7}$ | 1.8×10$^{-7}$ | 10.5 & 10.1 | 3.1×10$^6$ |
| SN8-300 | Si/SiN | 0.5×10$^{-3}$ | 1.3 | 1.7×10$^{-7}$ | 1.3×10$^{-7}$ | 10.4 & 8.7 | 2.1×10$^6$ |

**Graphical Abstract**

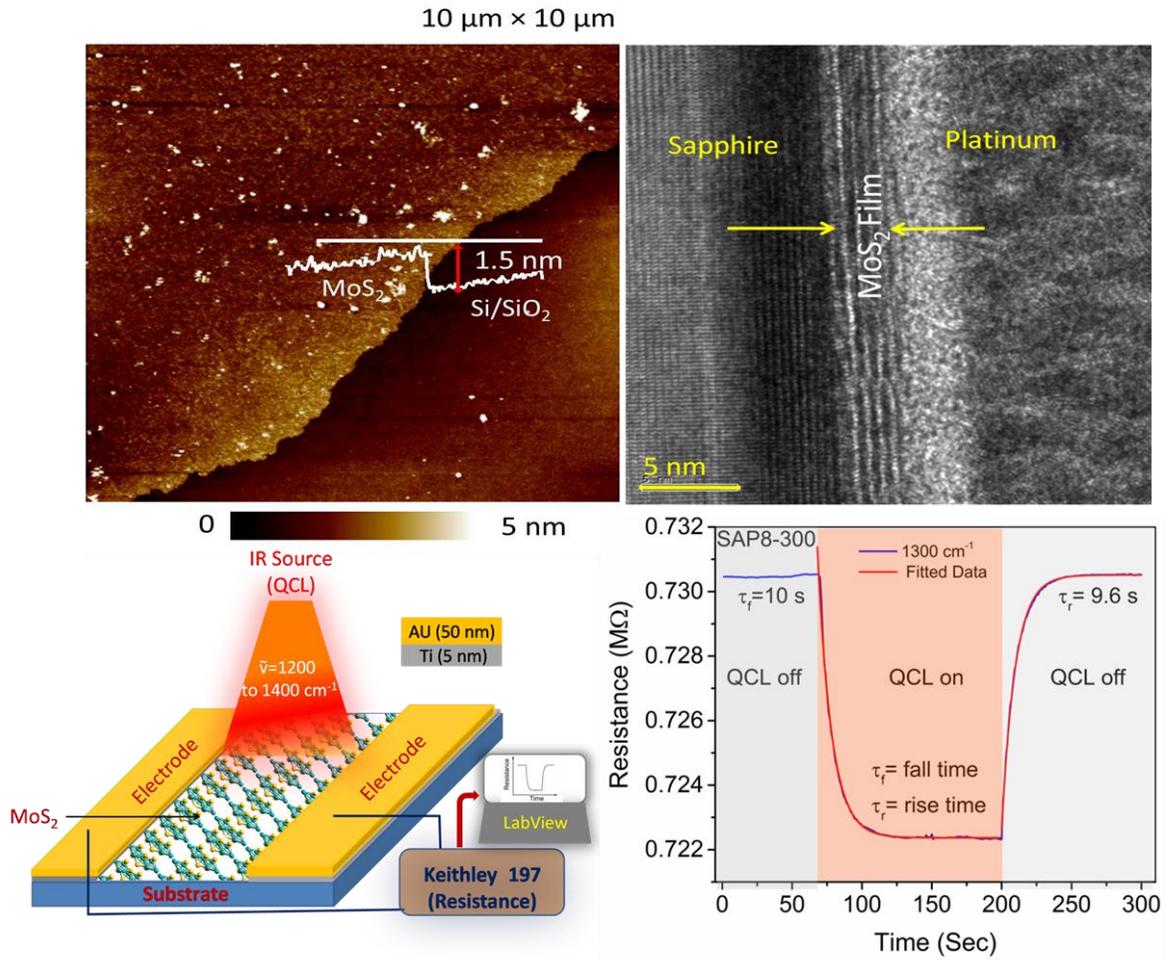

# Effect of interface on mid-infrared photothermal response of MoS$_2$ thin film grown by pulsed laser deposition

Ankur Goswami[a*], Priyesh Dhandaria[a], Soupitak Pal[b], M. Faheem Khan[a],
Željka Antić[a], Ravi Gaikwad[a], Kovur Prashanthi[a], Thomas Thundat[a*]

[a]Department of Chemical and Materials Engineering, University of Alberta, Edmonton, Canada
[b]Department of Chemical Engineering, University of California, Santa Barbara, USA

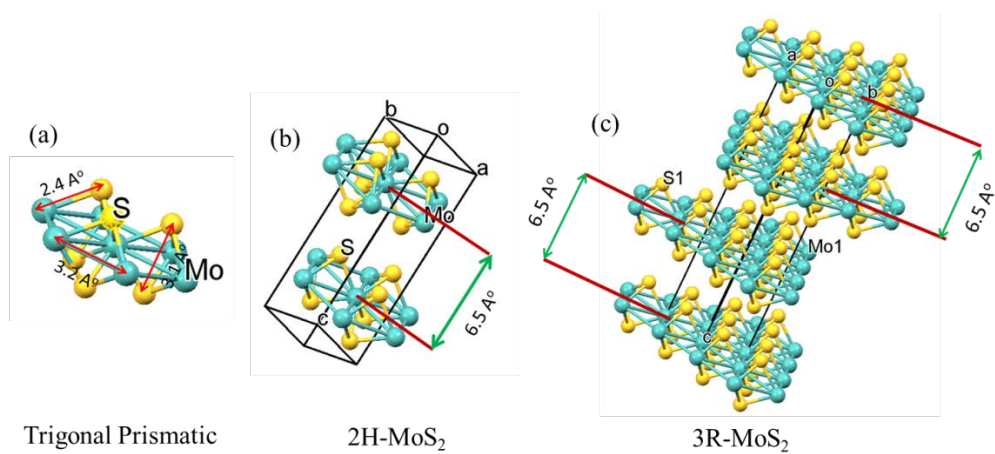

Scheme 1

Table-S1: Sample codes for MoS$_2$ on various substrates at different deposition conditions.

| Sample Code | Substrates | Deposition temperature (°C) | Laser Energy (mJ) | Time of Deposition (sec) |
|---|---|---|---|---|
| S7-20 | p-type Silicon (100) | 700 | 35 | 20 |
| S7-300 | p-type Silicon (100) | 700 | 35 | 300 |
| S8-20 | p-type Silicon (100) | 800 | 35 | 20 |
| S8-300 | p-type Silicon (100) | 800 | 35 | 300 |
| S8-5 | p-type Silicon (100) | 800 | 50 | 5 |
| SAP7-20 | Sapphire (0001) | 700 | 35 | 20 |
| SAP7-300 | Sapphire (0001) | 700 | 35 | 300 |
| SAP7-5 | Sapphire (0001) | 700 | 50 | 5 |
| SAP8-20 | Sapphire (0001) | 800 | 35 | 20 |
| SAP8-300 | Sapphire (0001) | 800 | 35 | 300 |
| SAP8-5 | Sapphire (0001) | 800 | 50 | 5 |
| SO7-20 | Thermal Oxide (Si/SiO$_2$) | 700 | 35 | 20 |
| SO7-300 | Thermal Oxide (Si/SiO$_2$) | 700 | 35 | 300 |
| SO8-20 | Thermal Oxide (Si/SiO$_2$) | 800 | 35 | 20 |
| SO8-300 | Thermal Oxide (Si/SiO$_2$) | 800 | 35 | 300 |
| SO8-5 | Thermal Oxide (Si/SiO$_2$) | 800 | 50 | 5 |
| SN7-20 | Silicon Nitride (Si/SiN) | 700 | 35 | 20 |
| SN7-300 | Silicon Nitride (Si/SiN) | 700 | 35 | 300 |
| SN8-20 | Silicon Nitride (Si/SiN) | 800 | 35 | 20 |
| SN8-300 | Silicon Nitride (Si/SiN) | 800 | 35 | 300 |
| SN8-5 | Silicon Nitride (Si/SiN) | 800 | 50 | 5 |

**Structural Characterization of MoS$_2$ deposited at 700 °C**

Fig. S1 shows the XRD of as-deposited film, grown on various substrates at 700 °C. It is observed that at 35 mJ laser energy, 20 s deposition time except SiN (SN7-20) none of the substrates show appreciable formation of MoS$_2$ film with sufficient crystallinity (Fig. S1a). At relatively longer time of deposition (300 s), thermal oxide of silicon (SO7-300) and SiN (SN7-300) substrates show appreciable growth of (0002) and a tiny peak of (103) MoS$_2$ as shown in Fig S1a. However, crystalline substrates (Si and sapphire) do not show considerable crystallinity even at higher time of deposition (300 s). In order to confirm that Raman spectroscopy were taken for all the samples.

There are two characteristic peaks of bulk MoS$_2$ (PLD target) normally observed at 379.7 and 404.8 cm$^{-1}$ as shown in Fig.S1b (see ESI).[1] These correspond to $E^1_{2g}$ and $A_{1g}$ vibration of 2H-MoS$_2$ Raman modes.[2] The former ($E^1_{2g}$) is the in-plane vibration of two opposite S atom corresponding to the Mo atom (in the x-y plane) while the latter ($A_{1g}$) corresponds to the out of plane vibration of only S atom in the normal plane (z-plane).[3] Generally for ultrathin (≤4 layers) MoS$_2$ the $E^1_{2g}$ band blue shifts whereas the $A_{1g}$ red shifts. From the difference between the Raman peaks frequencies (Δ) the number of stacked layer can be identified and listed in Table S1.[1]

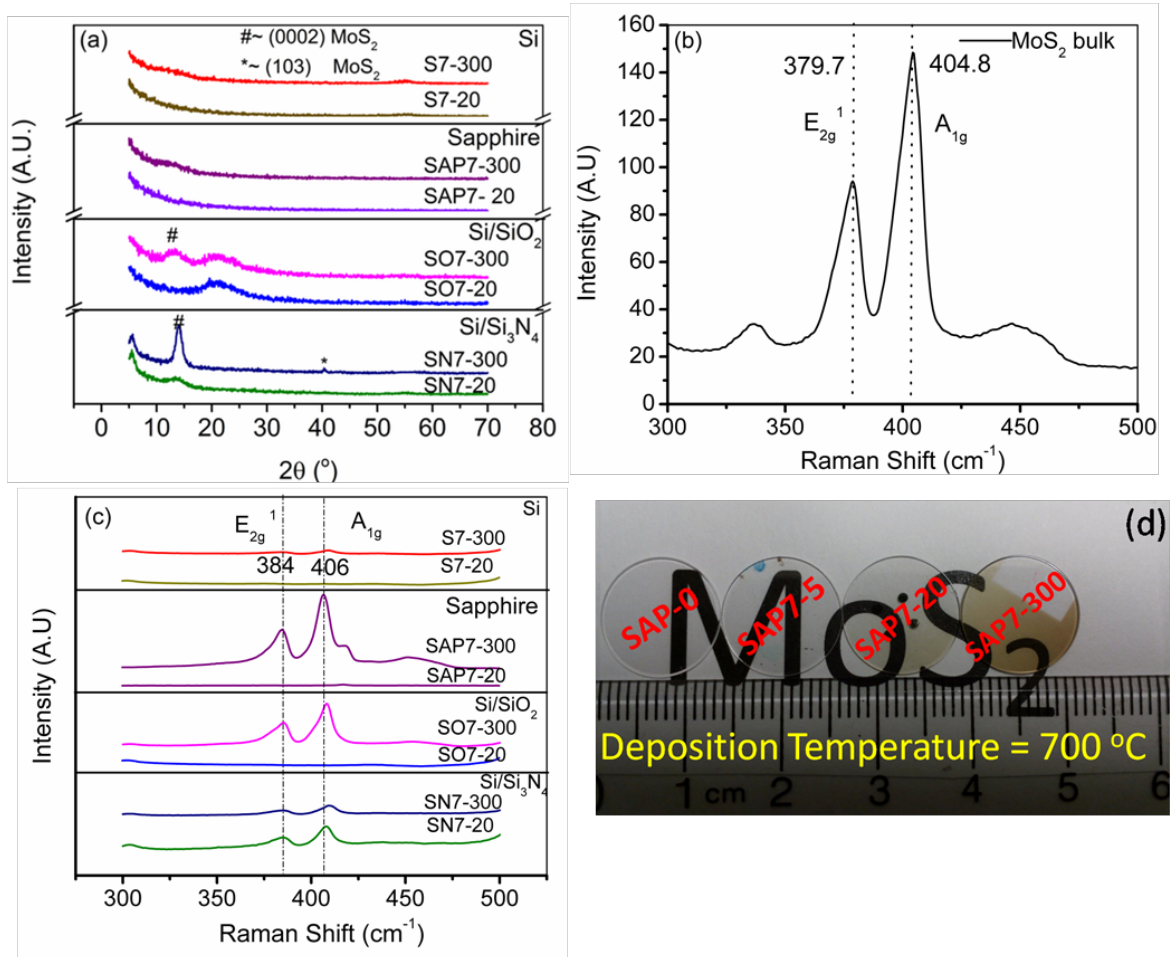

Fig. S1: (a) XRD of MoS$_2$ grown on different substrates at 700 $^o$C at 35 mJ laser energy. Raman spectra of MoS$_2$ (b) bulk, (c) grown on different substrates at 700 $^o$C at 35 mJ laser energy. (d) Photograph of the MoS$_2$ film grown on sapphire at 700 $^o$C different time and laser energy.

Fig.S1c shows the Raman spectra of as deposited film grown on various substrates at different time and temperature. It is observed from the Raman spectra (Fig. S1c) that at 700 $^o$C temperature and 20 s deposition time except SiN substrate no other substrates show any appreciable growth of MoS$_2$ (SN7-20). However, at 700 $^o$C temperature and 300 s of deposition time, the growth is similar for all the substrates and the deposited films show peak broadening.

Table S2: The $E^1_{2g}$ and $A_{1g}$ Raman peaks of as-deposited $MoS_2$ films on different substrates at various deposition time, temperature and laser energies.

| Samples | $E^1_{2g}$ (cm$^{-1}$) | $A_{1g}$ (cm$^{-1}$) | $A_{1g}-E^1_{2g}$ (cm$^{-1}$) ($\Delta\omega$) | Estimated Layer numbers |
|---|---|---|---|---|
| S7-20 | - | - | - | - |
| S7-300 | 384.3 | 408.5 | 25.2 | >5 or bulk |
| S8-20 | 384 | 407 | 23 | 3 to 4 |
| S8-300 | 382 | 408 | 26 | >5 or bulk |
| S8-5 | 385 (faint) | 407 (faint) | 23 | 3 to 4 |
| SAP7-20 | - | - | - | - |
| SAP7-300 | 384 | 406.3 | 22.3 | 3 to 4 |
| SAP8-20 | 383 | 407.8 | 24.8 | 4 to 5 |
| SAP8-300 | 382 | 408 | 26 | >5 or bulk |
| SAP8-5 | 385 | 407 | 22 | 2 to 3 |
| SO7-20 | - | - | - | - |
| SO7-300 | 385.6 | 408.5 | 23.1 | 3 to 4 |
| SO8-20 | 383.5 | 407.9 | 24.4 | 4 to 5 |
| SO8-300 | 384 | 409.6 | 25.6 | >5 or bulk |
| SO8-5 | 385.4 | 407.5 | 22.1 | 2 to 3 |
| SN7-20 | 384.8 | 407.7 | 22.9 | 2 to 3 |
| SN7-300 | 385.5 | 409.6 | 24.1 | 4 to 5 |
| SN8-20 | 384.7 | 408.9 | 24.2 | 4 to 5 |
| SN8-300 | 383.5 | 409.2 | 25.7 | >5 or bulk |
| SN8-5 | 385.4 | 408.3 | 22.9 | 2 to 3 |

Atomic force microscopy (AFM) was conducted to observe the evolution of the microstructure of MoS$_2$ film on different substrates. At 700 $^o$C temperature with laser energy of 35 mJ the film grown on the different substrates for 20 s of deposition time show grainy morphology as depicted in Fig. S2. This resembles to Volmer-Weber growth or island growth. The average roughness of all the films at these conditions is 4 to 6 nm which is relatively higher. But at higher deposition time (300 s) at same temperature (700 $^o$C) the growth morphology turns to be Stransky-Karstinov (SK) growth where both island and layer growth are observed.[4] At 700 $^o$C substrate temperature and 20 s of deposition time the MoS$_2$ species sit on top of each other because of the low atom mobility due to high adatom cohesive force. This dominates the surface adhesive force thereby increasing the surface roughness of the films.[5] However, as the temperature increased to 800 $^o$C while keeping deposition time constant (20 s), the roughness of the deposited films reduced because of the sufficient mobility of MoS$_2$ species on the all substrates. This leads to formation of smoother film similar to Frank-Vander Merwe (FM) growth where film grows in a layer-by-layer manner.

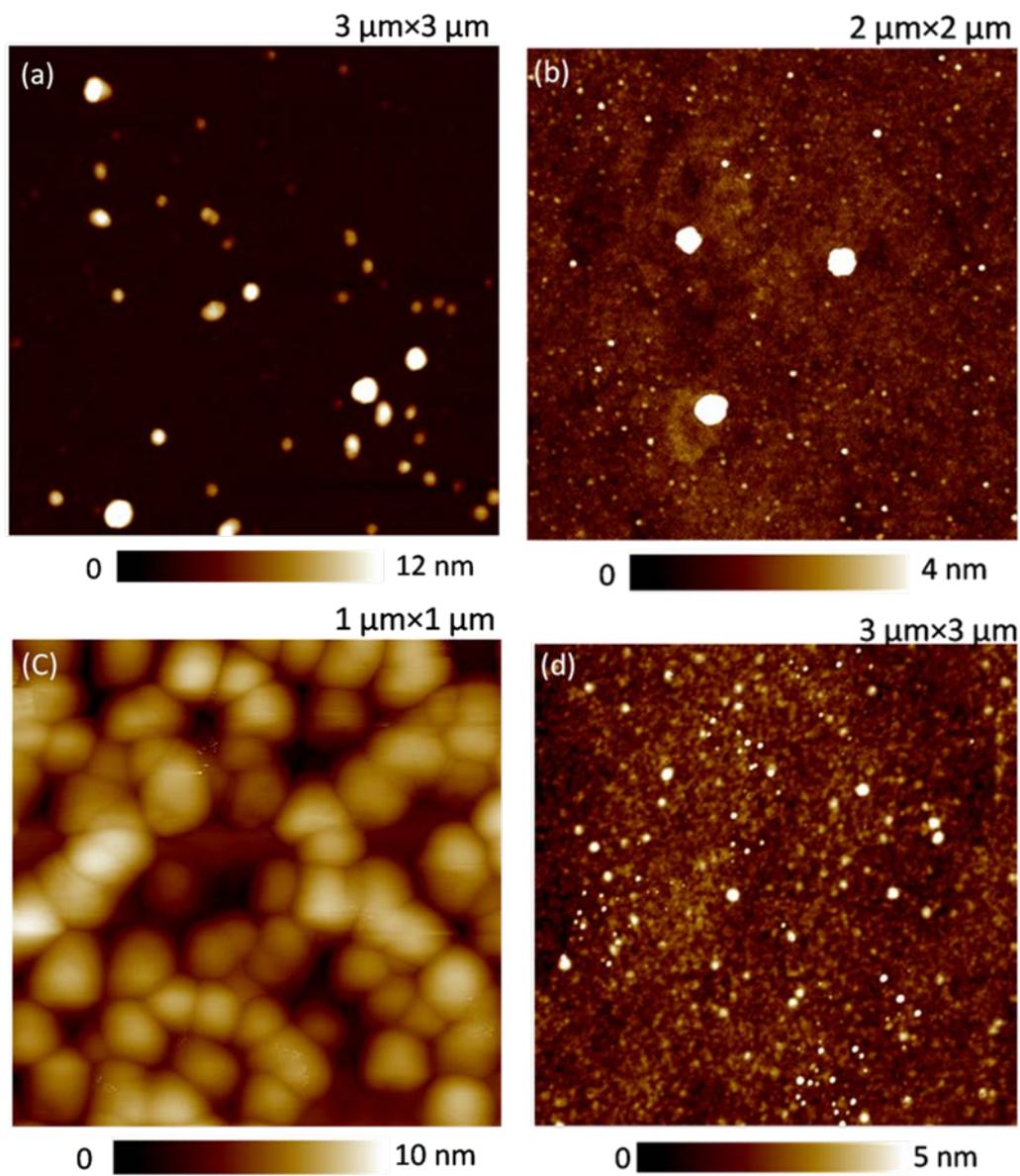

Fig. S2: AFM images of MoS$_2$ grown at 700 $^o$C on different substrates at 35 mJ laser energy for 20 s deposition times (a) silicon (S7-20) (b) sapphire (SAP 7-20) (c) thermally grown oxide SO7-20 (d) LPCVD grown SiN (SN7-20).

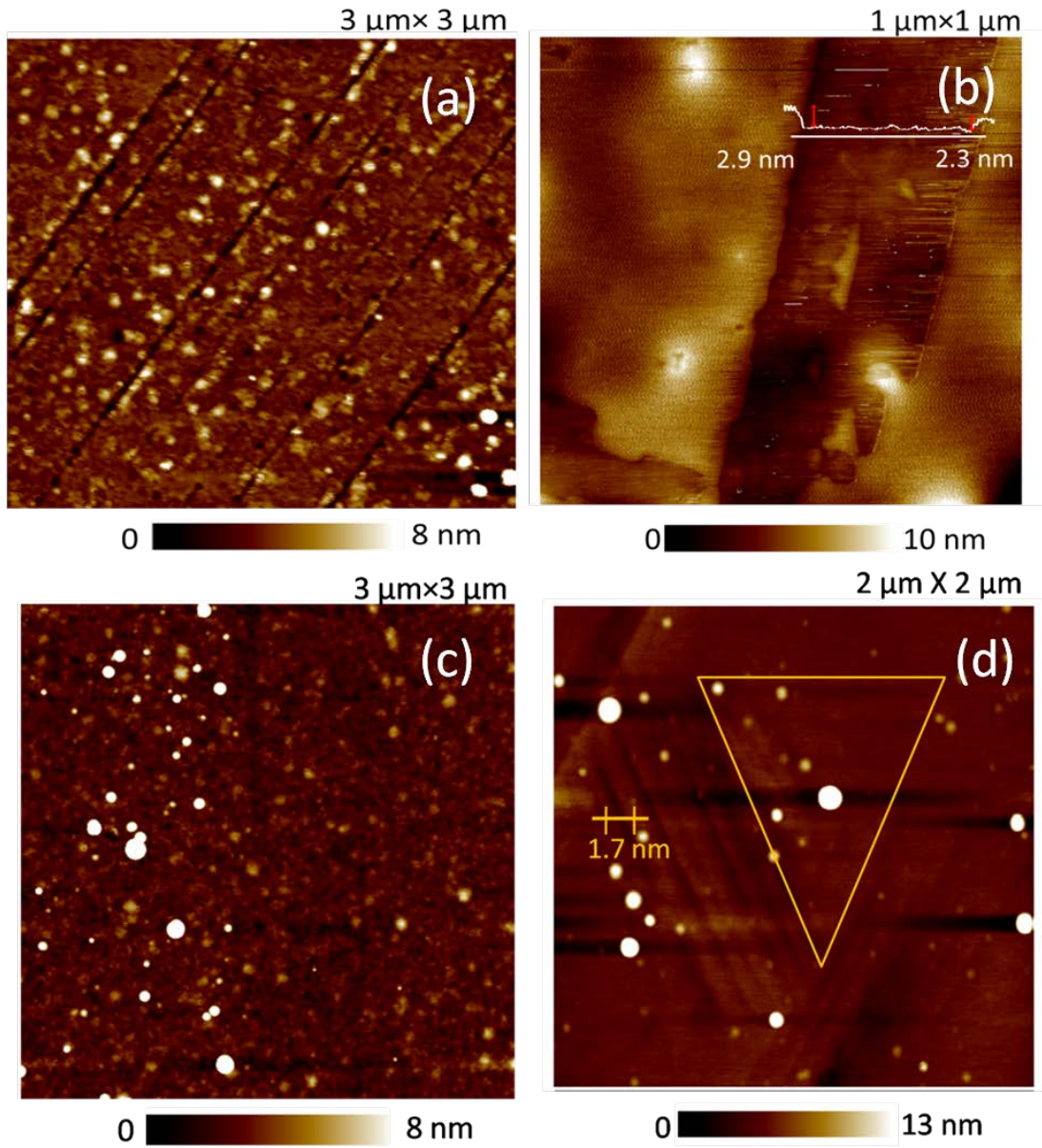

Fig. S3: AFM images of MoS$_2$ grown at 800 $^o$C on different substrates at 35 mJ laser energy for 20 s deposition times on (a) silicon (S8-20) (b) thermally grown oxide (SO8-20) (c) LPCVD grown SiN (SN8-20) and (d) MoS$_2$ grown at same temperature and energy for 300 s deposition time on sapphire (SAP8-300). Multiple layers have been grown with increasing the time.

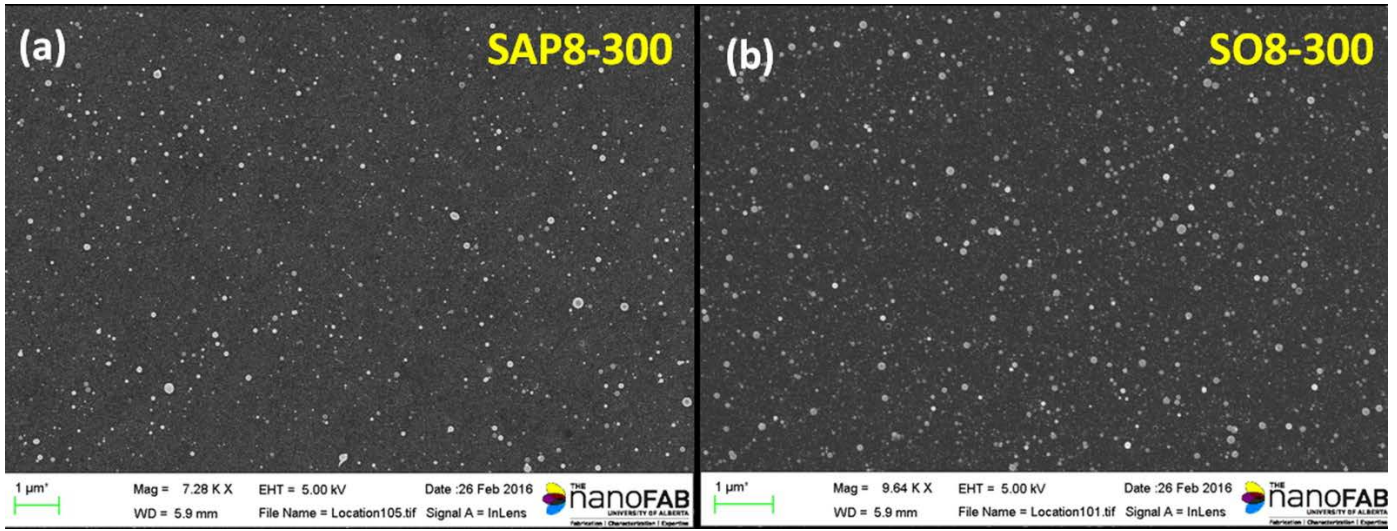

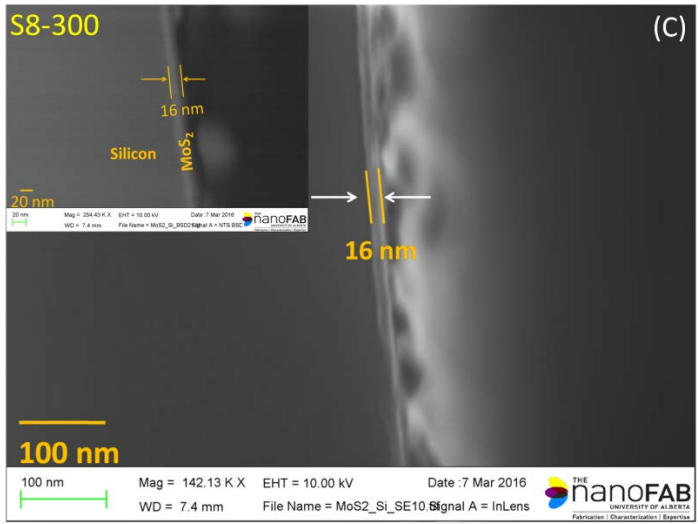

Fig. S4: SEM topography of MoS$_2$ deposited at 800 $^o$C for 300 s on (a) sapphire (SAP8-300) and (b) thermal oxide (SO8-300). (c) Cross sectional SEM of MoS$_2$ at same condition on silicon for thickness measurement (S8-300). Inset shows the back scattered image to distinguish the compositional contrast between silicon and MoS$_2$. All the other 300 s deposited films are of same thickness.

The oxidation states of all the MoS$_2$ samples deposited at 800 $^o$C for 20 s at different substrates were confirmed by high resolution XPS. The survey spectra of XPS of two samples (S8-20 and SAP8-20) are presented in Fig. S4a (see ESI). The XPS data (S8-20, SAP8-20, SO8-20 and SN8-20) are presented in Fig. S5 (see ESI). Two predominant peaks related to Mo$^{4+}$ are observed in every sample (as shown in S5 (a, c, e, g)) at 229 and 232.5 approximately. These two peaks correspond to Mo$^{4+}$ 3d$_{5/2}$ and 3d$_{3/2}$ respectively. In addition, all the spectra show S 2s peaks at around 226.3 to 226.7 eV. These peaks correspond to Mo-3d of 2H structure. However, two unwanted peaks also appeared around 233 and 236 eV which corresponds to Mo$^{6+}$3d$_{5/2}$ and Mo$^{6+}$3d$_{3/2}$. These two peaks signify the possibility of MoO$_3$ existence which may be due to contamination or due to oxidation from long time exposure of MoS$_2$ to the outside atmosphere prior to XPS experiments. Nevertheless, all the samples show S 2p doublet (as shown in Fig. S5b, S5d, S5f, S5h) (see ESI) around 161.9 to 162.5 eV (S 2p$_{3/2}$) and 163.5 eV (S 2p$_{1/2}$). There is no other amorphous sulphur observed in any of the samples which confirms all the samples are of crystalline quality. Further, it is also observed from atomic fraction calculation of XPS data (using Casa-XPS software) that Mo and S ratio lied between 1:1.9 to 1:1.95 which confirms MoS$_2$ grown by PLD were nearly stoichiometric.

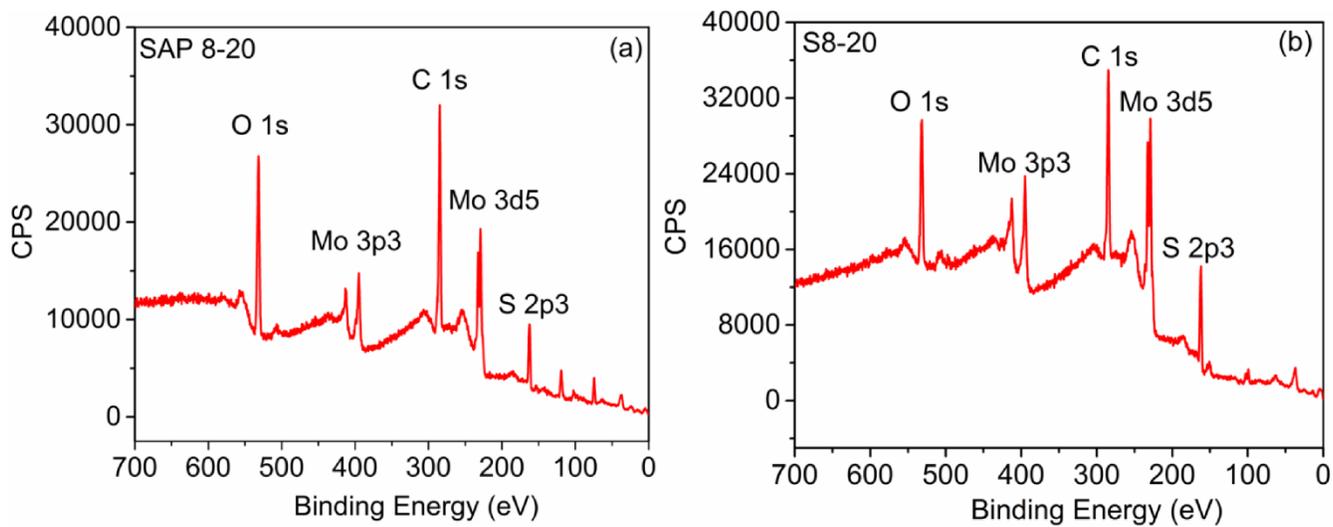

Fig. S5: XPS survey spectrum of $MoS_2$ on (a) sapphire (SAP8-20) (b) silicon (S8-20).

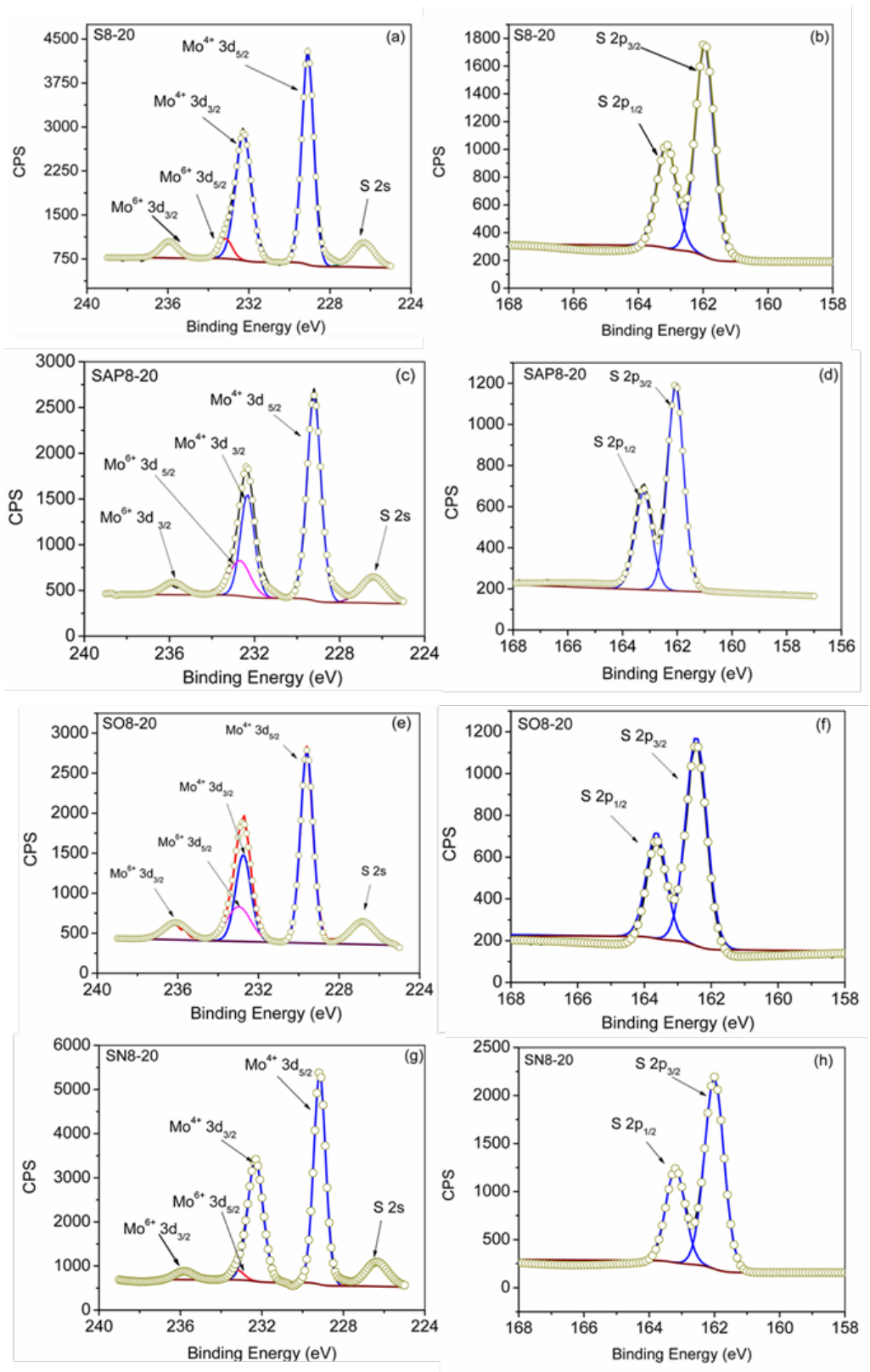

Fig. S6: XPS spectra of MoS$_2$ on Si , SiO$_2$, SiN showing (a),(c),(e) Mo 3d, and (b),(d), (f) S 2s and S 2p core level peak regions respectively.

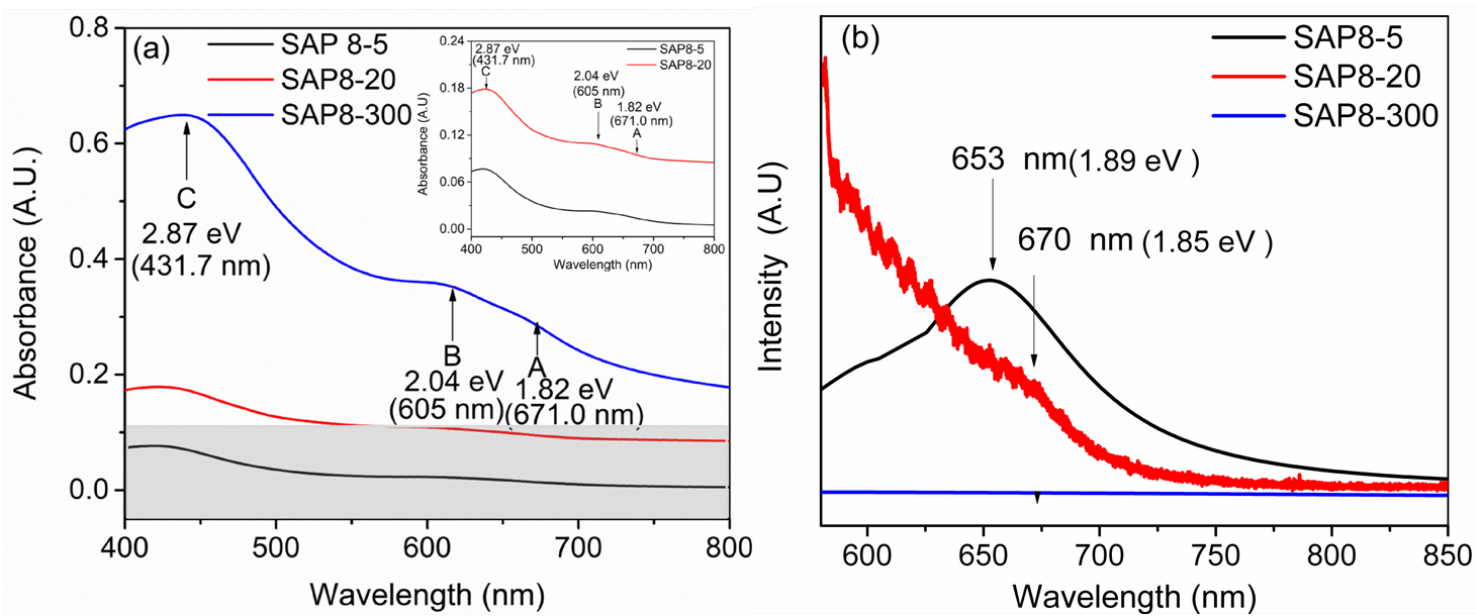

Fig. S7: Optical and photoluminescence spectra of $MoS_2$ grown on sapphire at 800 $^o$C at 35 mJ energy for different times.

$MoS_2$ grown on sapphire has been chosen for UV-Vis study due to an excellent optical transparency of sapphire. Fig. S6a shows the optical absorbance spectra of $MoS_2$ grown on sapphire at 800 $^o$C at three different times (5 s, 20 s and 300 s). It is observed from the spectrum that there are two clearly well-known excitonic absorption bands "A" and "B" appeared at 671 nm (1.82 eV) and 605 nm (2.04 eV) respectively for SAP8-300.[6] In case of SAP8-5 and SAP8-20, "A" band is relatively feeble but "B" band is significant (in the inset). However, both the bands ("A" and "B") correspond to the band-edge excitons which happened due to the excitonic transitions at the Brillouin zone K point.[7] Additionally, peak ("C") appeared at lower wavelengths at 431 nm which corresponds to the van Hove singularities in the electronic density of states in the layered $MoS_2$ which is common in 2D layered materials.[8]

Photoluminscence (PL) spectra of the same samples are shown in Fig. S6b. The peak at 653 nm (1.89 eV) of SAP8-5 sample shows the direct excitonic transition from the band gap which contains 2 to 3 layers as confirmed from Raman (tableS1) and AFM. As the layer number increases the peak red shifted and the strength of the intensity reduces. For bulk $MoS_2$ (SAP8-300) the signal intensity is near negligible as reported in earlier literature.[9] This confirms the band gap of $MoS_2$ changes with layer numbers and we believe the band gap of the multilayer $MoS_2$ deposited here in all the substrates converges to 1.29 eV as reported in literature.[10]

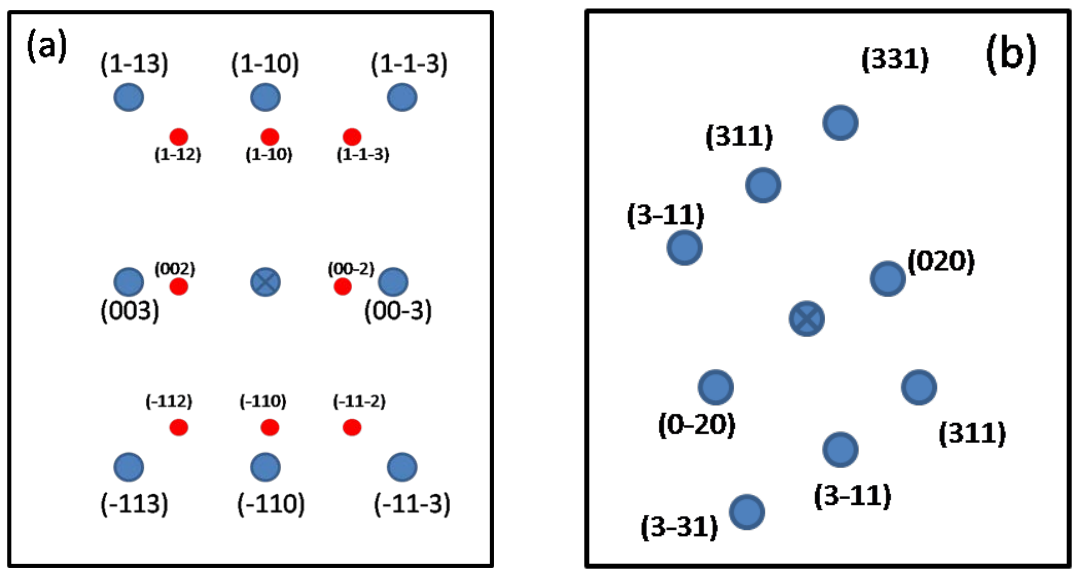

Fig. S8: (a) and (b) Orientation relationship of MoS$_2$ on sapphire from FFT pattern of TEM cross-sectional image. In Fig 6 (a) blue and red dots represent FFT of sapphire substrate and MoS$_2$ respectively. In Fig 6(b) no MoS$_2$ spots are seen since the film was not in edge-on condition.

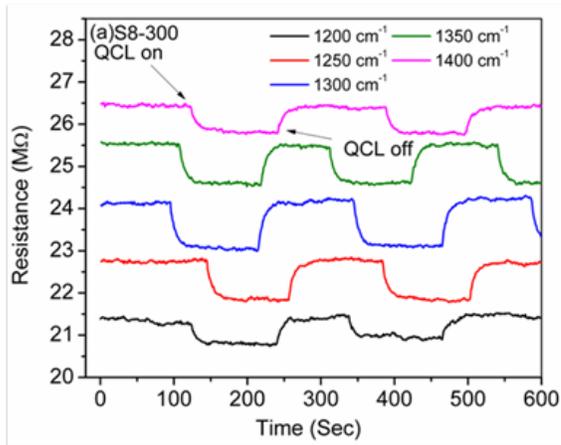
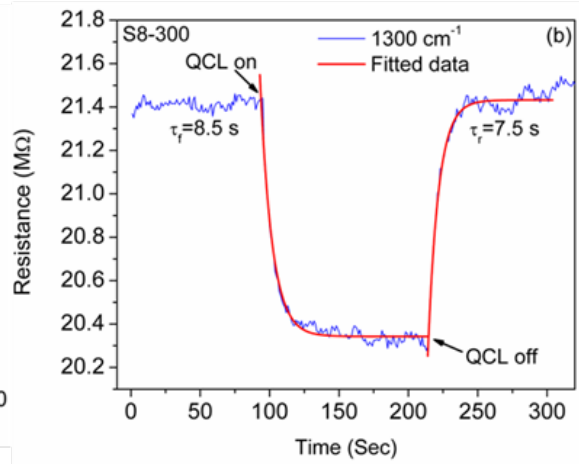
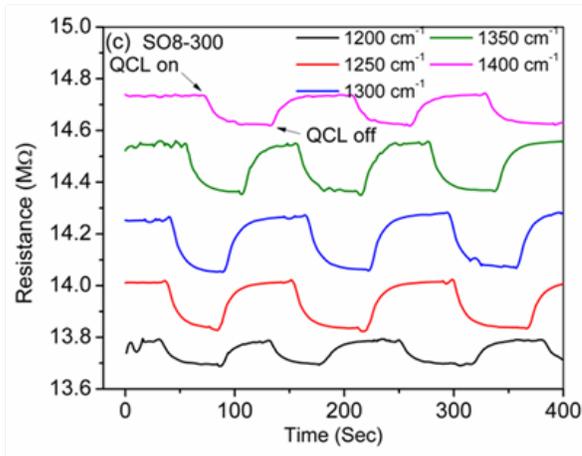
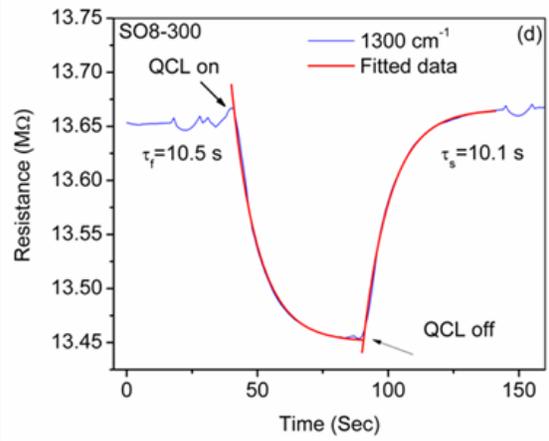
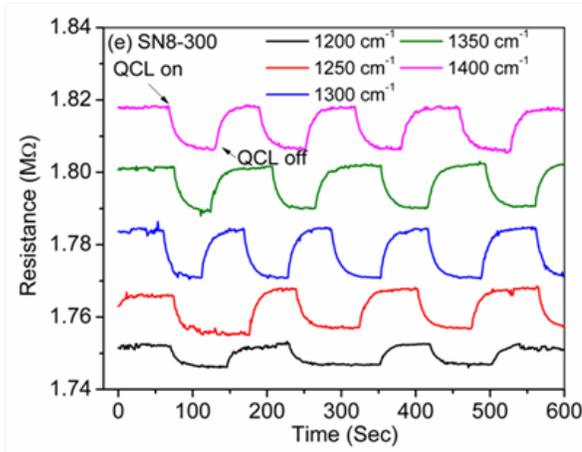
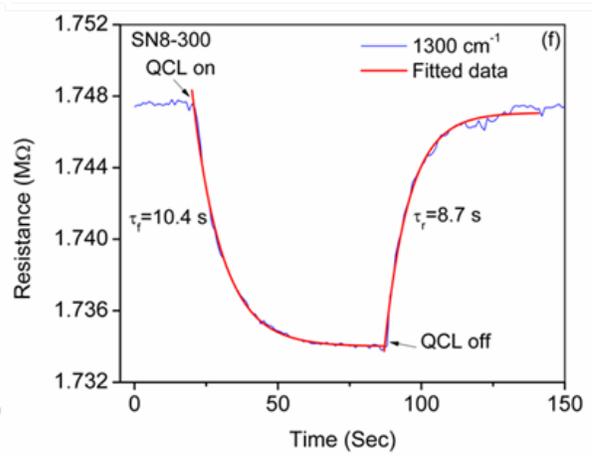

Fig S9: Variation of resistance and the response time of MoS$_2$ on different substrates under IR illumination.

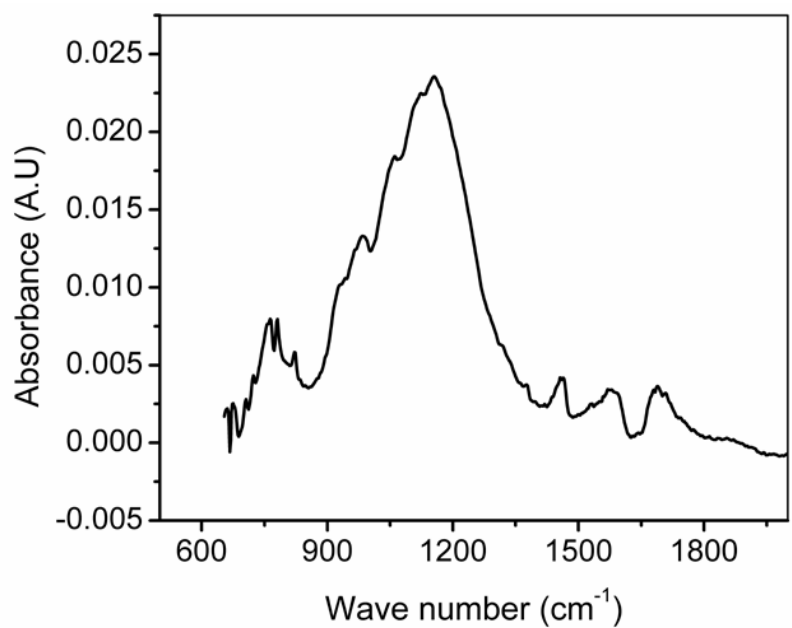

Fig. S10: FTIR spectra of bulk MoS$_2$ in order to show the broadband IR absorption from 600 to 2000 cm$^{-1}$ wave number.

**Responsivity calculation:**

Responsivity of any detector is measured by the voltage or current generation with respect to the incident radiant power falls on the detector.[11] Here, MoS$_2$ film acts as a detector which changes its resistance upon incident IR without any external bias. The equivalent circuit of the detector with the multimeter unit is shown in Fig. S11 (see ESI). Multimeter feeds constant voltage ($V_{in}$ = 4.3 V) to the circuit which goes through an internal circuit resistance. The internal resistance varies from 1 MΩ to 10 MΩ depending on the range of resistance needs to measure. Hence, the whole circuit works as a voltage divider. Therefore the voltage drop due to the resistance change in response to the IR radiation can be calculated from the following formula.

$$\Delta V = \frac{\Delta R_{IR(MoS_2)}}{R_M + R_{0(MoS_2)}} V_{in} \tag{S1}$$

Where $\Delta V$ is the voltage drop because of the resistance change of the MoS$_2$ resister due to IR radiation. $R_M$ is the multimeter's internal resistance which varies from 1 MΩ to 10 MΩ depending on the range of resistance required to measure. $R_{0(MoS_2)}$ is the initial resistance of the MoS$_2$ film before the IR radiation. $\Delta R_{IR(MoS2)}$ is the change of resistance after IR irradiation. Hence, the responsivity is $\Delta V / W$ where W is the power of incident radiation. Here, we measured the responsivity at highest average power of QCL at 1300 cm$^{-1}$ (7.7 µm) i.e. 25 mW.

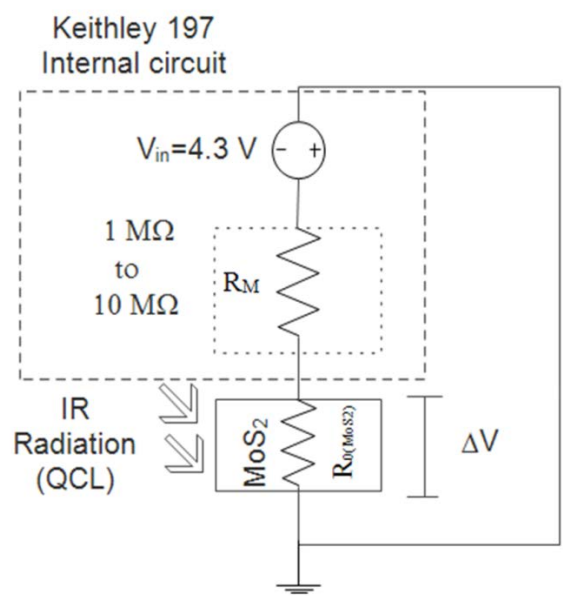

Fig. S11: Equivalent circuit to calculate responsivity of MoS$_2$ thin film.